\documentclass[12pt]{iopart}
\usepackage{graphicx}
\usepackage{color}
\usepackage{amssymb}
\usepackage{comment}
\usepackage{float}
\usepackage{setspace}
\usepackage{tikz}
\usetikzlibrary{shapes,arrows}
\usepackage{color}
\usepackage[normalem]{ulem}

\definecolor{scarred}{rgb}{0.75,0.0,0.0}
\usepackage{hyperref}
\usepackage{subcaption}

\hypersetup{
colorlinks=true,final=true,
        linkcolor=red,
        citecolor=blue,
        filecolor=blue,
        urlcolor=blue,
}

\begin{document}
\title{Interplay of strong correlations and covalency in ionic band insulators}
\author{Nagamalleswararao Dasari$^1$, Sujan K. K.$^1$, Juana Moreno$^2$, and N. S. Vidhyadhiraja$^1$}
\address{$^1$ Jawaharlal Nehru Centre For Advanced Scientific Research, Jakkur, Bangalore 560064, India.}
\address{$^2$ Department of Physics $\&$ Astronomy, Louisiana State University, Baton Rouge, LA 70803-4001, USA.}

\begin{abstract}
We address the role of electronic correlations in different kinds of band insulators by using the two-orbital Hubbard model within the dynamical mean-field theory (DMFT). An intriguing finding is that electronic correlations turn a metal into a band insulator when ionicity and covalency are equal in ratio.
We conclude that the electronic correlations favour metallicity when the covalency is smaller than the ionicity, while they favour insulating behaviour when the covalency is greater than ionicity.  
\end{abstract}

\maketitle

\section{Introduction}
The recent discovery of interaction-driven topological phases\cite{turner2013beyond,chen2013symmetry,budich2013fluctuation,amaricci2015first}, such as fractional quantum-Hall states, spin-liquids, Kondo-insulators and bosonic topological phases, has created a huge interest in, otherwise considered to be mundane, band insulators. Some questions of fundamental interest in band insulators are: how do correlations drive a band insulator into a metal? and a Mott insulator(MI)?, and are correlated band insulators fundamentally different from simple band insulators with identical charge and spin excitation gaps? These issues have been addressed theoretically in all dimensions, from one to infinity, by studies of model Hamiltonians such as the ionic Hubbard model\cite{kubler2003engineering,manmana2004quantum,batista2004exact,garg2006can,kancharla2007correlated,craco2008electronic,paris2007quantum,byczuk2009insulating,kim2014finite,hoang2010metal}, a two-sublattice model with inter-orbital hybridization\cite{sentef2009correlations,euverte2013magnetic}, a two-band Hubbard model with crystal field splitting\cite{werner2007high} and a bilayer model with two identical Hubbard planes\cite{moeller1999rkky,fuhrmann2006mott,kancharla2007band,fabrizio2007gutzwiller,hafermann2009metal}. 

The ionic Hubbard model, which comprises a two-sublattice system having orbital energies $V$ and $-V$ with a local Coulomb repulsion, drew a lot of attention after the pioneering work by Arti Garg et. al.,\cite{garg2006can}, which showed that correlations can turn a band insulator into a metal and for higher interaction strengths, $U$, into a Mott insulator. The $U-V$ phase diagram, found through an iterative perturbation theory (IPT) solution of the self-consistent impurity problem within dynamical mean field theory (DMFT), exhibited a finite metallic region, which transformed into a line
at large $U$ and $V$, as should be the case in the exactly known atomic limit. Later studies using a modified form of IPT~\cite{craco2008electronic}, numerical renormalization group at zero temperature \cite{byczuk2009insulating}, and the continuous time quantum Monte-Carlo (CTQMC) method\cite{kim2014finite}, while confirming the existence of an intervening metallic phase, were not in agreement about the extent of the metallic region. Furthermore, one could ask if there exist parameters other than interaction strength that could induce metallicity in band insulators, and what would be the interplay of interactions with such an athermal parameter. In this work, we obtain the results from CTQMC, while also answering the latter question within a two orbital Hubbard model with on-site repulsion, $U$, between electrons of opposite spin. The novelty of our model is embodied by a parameter, $x \in [0,1]$, which may be interpreted as the degree of ionicity, while $1-x$ is concomitantly interpreted as the degree of covalency. Such a parametrization permits us to explore the interplay of ionicity and covalency in interacting band insulators. So for $x=1$, we obtain purely ionic band insulators\cite{garg2006can} while for $x=0$, the model reduces to purely covalent insulators\cite{sentef2009correlations}.
An investigation of correlations in polar-covalent insulators is important in its own right. The characteristic charge gap in these insulators is of the order of a few meV and given by inter-orbital hybridization between partially filled bands\cite{sentef2009correlations,kunevs2008temperature}. The canonical example of covalent insulators are FeSi and FeSb$_2$\cite{schlesinger1993unconventional,petrovic2005kondo}. The temperature evolution of the charge gap in these systems is such that it closes at a temperature which is low relative to the size of the $T=0$ gap. Additionally, a temperature-dependent spectral weight transfer to high frequencies ($\approx$1 eV) above the gap edge is seen in the optical conductivity. 

One of our main findings is that, while the two extremes of $x=0$ and $x=1$  are indeed band insulators, albeit of different kinds, the $x=0.5$ case turns out to be a metal even in the non-interacting case. Further, the metal at $U=0$ turns into a correlated band insulator even for infinitesimal interactions, and a re-entrant metallic phase is found at higher interactions, beyond which a Mott insulator is obtained. We find a rich phase diagram in the $U-T$ plane that is strongly dependent on the degree of covalency (or ionicity).  

The paper is organized as follows: In Sec.~\ref{Model5.1}, we define the model and methods chosen to study correlation effects in different kinds of band insulators. 
In Sec .~\ref{Model5.2}, we present and discuss our numerical results.
Finally, in Sec.~\ref{Model5.3}, we present our conclusions.

\section{Models and Methods}
\label{Model5.1}

We consider a two-orbital Hubbard model with both orbitals having a local Coulomb interaction between electrons of opposite spin on the same orbital. In the second quantized notation, the Hamiltonian reads,
\begin{equation}
\hspace*{-\mathindent}\,\,\,
\begin{array}{c}
\displaystyle {\cal H} = \sum_{k\sigma} 
\left(
\begin{array}{cc}
 c^\dagger_{k 1 \sigma} & c^\dagger_{k 2 \sigma}
\end{array}
\right)
\mathbf{H}_{\sigma}(k)
\left(
\begin{array}{c}
  c^{\phantom\dagger}_{k 1 \sigma} \\[6pt] 
  c^{\phantom\dagger}_{k 2 \sigma}
\end{array}
\right) 
 - \mu \sum_{i\alpha\sigma} \hat{n}_{i\alpha\sigma}
 + \sum_{i\alpha\sigma} \frac{U}{2}\,\hat{n}_{i\alpha\sigma}\hat{n}_{i\alpha\bar{\sigma}} \,,
\end{array}
\label{eq:Ham}
\end{equation}
\begin{equation}
\hspace*{-\mathindent}\,\,\,
\begin{array}{c}
\displaystyle \mathbf{H}_{\sigma}(\mathbf{k},x) 
   = x\,\mathbf{H}_{IBI} + (1-x)\,\mathbf{H}_{CI}  = x\left(
\begin{array}{cc}
\Delta & \epsilon_{k\sigma} \\
\epsilon_{k\sigma} & -\Delta
\end{array}
\right)
+
(1-x)\left(
\begin{array}{cc} 
\tilde{\epsilon}_{k\sigma} & V \\
V & -\tilde{\epsilon}_{k\sigma}
\end{array}
\right)\,
\end{array}
\label{eq:matrix}
\end{equation}
where $c^{\dagger}_{i\alpha\sigma} (c_{i\alpha\sigma})$ creates (annihilates) an electron at lattice site $i$, in orbital $\alpha=1,2$ with spin $\sigma \in \{\uparrow, \downarrow \}$. 
We set the chemical potential $\mu$ = $\frac{U}{2}$ so that each unit cell is half filled with an average occupancy of 2. The unit cell thus consists of two orbitals $1$ and $2$. The parameter $x$ interpolates between IBI and CI.
Here IBI corresponds to $x=1\,(H_\sigma(\mathbf{k},1)=H_{IBI})$, while the CI is obtained at $x=0\,(H_\sigma(\mathbf{k},0)=H_{CI})$, hence $x$ represents the fraction of ionicity, while $1-x$ represents covalency. In the IBI, a two-orbital system has a staggered ionic potential $\pm\Delta$ across orbitals 1 and 2, along with a momentum-dependent hybridization ($\epsilon_{k\sigma}$) between them. The CI is characterized by two
bands having opposite sign of the hopping parameter and a $k$-independent hybridization $V$. The diagonal dispersion in the CI corresponds to intra-band electron hopping, while the off-diagonal dispersion in the IBI corresponds to inter-band electron hopping. By varying the parameter $x$ from 1 to 0, we interpolate smoothly between a purely ionic limit (for $x=1$) and a purely covalent limit($x=0$). In other words, the percentage of covalency in the ionic band insulator increases as we decrease $x$ from $1$ to $0$. 

The motivation to build and study the above Hamiltonian is twofold: (a) There are three primary chemical bonds namely ionic, covalent and metallic. But in practice, a perfect ionic bond does not exist,
and quantifying the covalency or the ionicity of a given bond is not without ambiguities\cite{kittel2018kittel,ashcroft1976solid}. Depending upon the percentage of covalency, the properties of the system changes drastically\cite{kittel2018kittel,ashcroft1976solid}. Eq.~(\ref{eq:matrix}) is one the simplest, and of course non-unique, ways of parametrizing a system wherein the bonding has both ionic and covalent character. (b) Another perspective from the view point of real materials is that the non-interacting Hamiltonian $\mathbf{{H}_{\sigma}(k)}$ could have both inter-unit cell and intra-unit cell hybridizations, where inter-unit cell hopping is often neglected in model calculations\cite{han1998multiorbital}.

Throughout the paper, we consider the case where V = $\Delta$, $\epsilon_k$ = $\tilde{\epsilon}_k$, and a semicircular density of states. Although these are specific parameter choices, the results we obtain are quite  general and applicable to other choices. We are mainly interested in local single particle electron dynamics, which is given by the momentum sum of the lattice Green's function,
\vspace{0.5em}
\begin{equation}
\qquad\mathbf{G}_{\sigma}(\omega^+) = 
\sum_{{\mathbf{k}}} \left[(\omega^+ + \mu)\mathbb{I} -
 \mathbf{{H}_{\sigma}(k)} - \mathbf{{\Sigma}_{\sigma}}(\omega^+)\right]^{-1}\,,
\label{eq:gloc}
\end{equation} 
\vspace{0.5em}
where $\omega^+ = \omega+i\eta$ and $\eta\rightarrow
0^+$, and $\mathbb{I}$ is the identity matrix. We calculate the local single particle propagators within the DMFT framework, wherein the single particle
irreducible self-energy $\mathbf{\Sigma}_{\sigma}(\omega^+)$ is local, and will be determined by solving the auxiliary Anderson impurity model. The local, interacting Green's function (Eq.~(\ref{eq:gloc})) may be related to the non-interacting Green's function ${\bf{G}}_{0\sigma}(\omega^+)$ through the Dyson equation: 
\vspace{0.5em}
\begin{equation}
\qquad\qquad\qquad {\bf{G}}^{-1}_{0,\sigma}(\omega^+) = {\bf{G}}^{-1}_{\sigma}(\omega^+)+{\bf{\Sigma}}_{\sigma}(\omega^+) \,.
\end{equation}
\vspace{0.5em} 
The structure of $\mathbf{{H}_{\sigma}(k,\mathnormal{x})}$ determines the form of the impurity Green's functions, which for orbital (or sublattice) 1 is given by,
\begin{equation}
\qquad G_{1\sigma}(\omega^{+}) = \int d\epsilon \frac{ \zeta_{2\sigma}(\omega^{+},\epsilon)  \rho_0(\epsilon)}{\zeta_{1\sigma}(\omega^{+},\epsilon) \zeta_{2\sigma}(\omega^{+},\epsilon) - [V(1-x)+\epsilon x]^2}\,, 
\label{eq:green}
\end{equation}
where
\begin{equation*}
\qquad\begin{array}{c}
\zeta_{1\sigma}(\omega^{+},\epsilon) = 
  \omega + i\eta + \mu - [Vx + \epsilon(1-x)]
  - \Sigma_{1\sigma}(\omega^{+}) \,, \\[1ex]
\zeta_{2\sigma}(\omega^{+},\epsilon) = 
  \omega + i\eta + \mu + [Vx + \epsilon(1-x)]
  - \Sigma_{2\sigma}(\omega^{+}) \,,
\end{array}
\end{equation*}
and $\rho_0(\epsilon)$ = $\frac{2}{\pi D}$ $\sqrt{1-(\epsilon/D)^2}$. $D$ is our energy unit and $\eta\rightarrow
0^+$ is the convergence factor. In this work, we fix $D=1$ unless otherwise specified. At half-filling the Hamiltonian has particle-hole symmetry between orbitals and is unchanged under 
the following transformation:
\vspace{0.5em}
\begin{equation*}
\qquad\qquad\qquad\begin{array}{c}
c^{\dagger}_{i 1 \sigma} = d_{i 2 \sigma}, \quad\quad
  c^{\dagger}_{i 2 \sigma} = - d_{i 1 \sigma} \, .
\end{array}
\end{equation*}
\vspace{0.5em}
The impurity Green's functions of both orbital are then related as
\vspace{0.5em}
\begin{equation*}
\qquad\qquad\qquad G_{1\sigma}(\omega^{+}) = - \left[G_{2\sigma}(-\omega^{+})\right]^{*}\,. \\
\end{equation*}
\vspace{0.5em}
By using the above symmetry relation, we can readily show that the self-energy and $\zeta$ terms fulfill the following relations,
\vspace{0.5em}
\begin{equation*}
\qquad\qquad\qquad\begin{array}{c}
\Sigma_{1\sigma}(\omega^{+}) = U - \left[\Sigma_{2\sigma}(-\omega^{+})\right]^{*}, \\[1ex]
\,\zeta_{1\sigma}(\omega^{+},\epsilon) = -\left[\zeta_{2\sigma}(-\omega^{+},\epsilon)\right]^{*}\,.
\end{array}
\end{equation*}
\vspace{0.5em}
Then equation~(\ref{eq:green}) can be written as,
\vspace{0.5em}
\begin{equation}
G_{1\sigma}(\omega^{+})=\int d\epsilon \frac{\zeta^*_{1\sigma}(-\omega^{+},\epsilon) \rho_0(\epsilon)}{\zeta_{1\sigma}(\omega^{+},\epsilon) \zeta^*_{1\sigma}(-\omega^{+},\epsilon) + [V(1-x)+\epsilon x]^2}\,. 
\label{eq:general}
\end{equation}
\vspace{0.5em}
Now we are going to present a few analytical results for the density of states (DOS) at the Fermi level ($\omega=0$) and subsequently, we will discuss our numerical results.  

\section{Numerical results and Discussion}
\label{Model5.2}

In the following, we present our numerical results of Eq.~(\ref{eq:Ham}) at finite temperature. Unless otherwise stated, we fix $k$ independent hybridization $V$ is 0.5. We solve the auxiliary Anderson impurity model of Eq.~(\ref{eq:Ham}) within DMFT. As an impurity solver, we use the hybridization expansion continuous-time quantum Monte-Carlo (HY-CTQMC)\cite{werner2006continuous,bauer2011alps}. In this work, we obtain the single particle Green's function and self-energy on the Mastubara axis 
and analyzed these functions without carrying out analytical continuation on the real-frequency axis. 

\subsubsection{Ionic band insulator ($x$=1):}

\begin{figure}[h!]
    \centering
    \begin{subfigure}[b]{0.49\textwidth}
        \centering
        \includegraphics[scale=0.59,trim={0 0 0 1}, 
        clip]{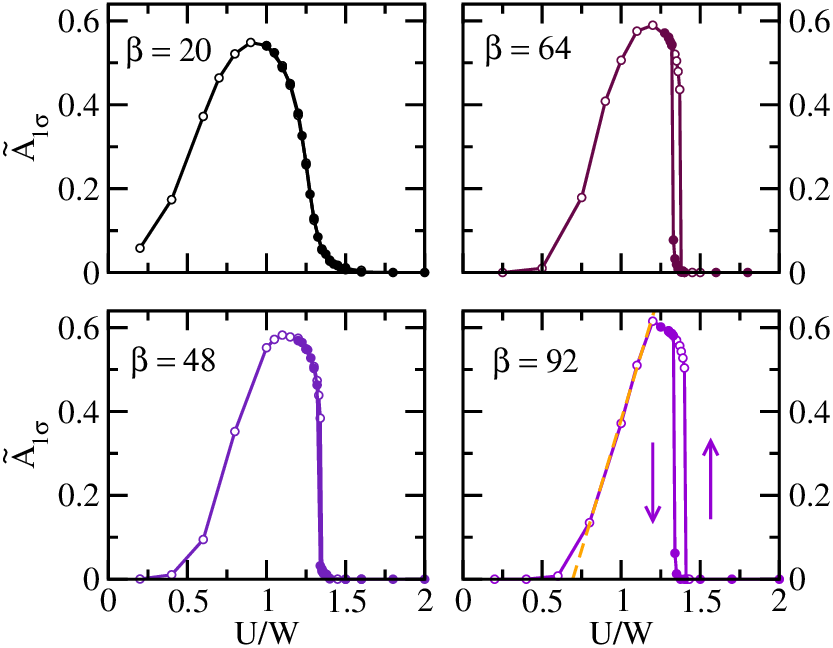}
        \caption{}
        \label{fig:x1_aw0}
    \end{subfigure}
    \hfill
    \begin{subfigure}[b]{0.49\textwidth}
        \centering
         \includegraphics[scale=0.5,trim={0 0 0 1}, 
        clip]{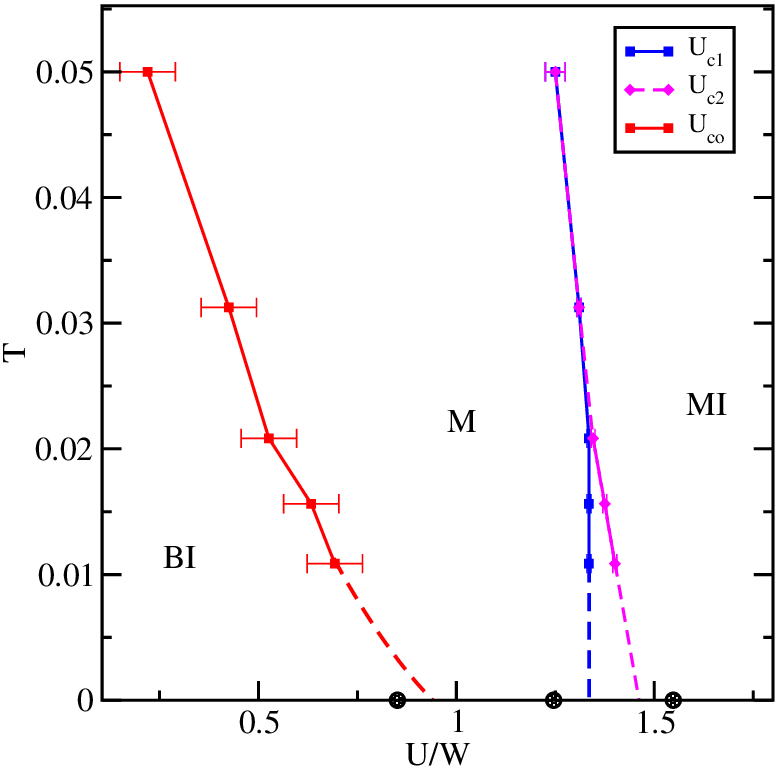}
        \caption{}
        \label{fig:x1_phase}
    \end{subfigure}
    \caption{(Color online) a) Fermi-level spectral weight $\tilde{A}_{1\sigma}$ as a function of U for different $\beta$ values obtained from HY-CTQMC for $x=1$. Open circles correspond to increasing $U$, while closed circles represent decreasing $U$. To find the crossover of BI-M, we fit the linear region as shown in the data of $\beta=92$(The expression of the fit(orange dashed line) is $y=-0.86+1.24x$). The purple arrow(for $\beta=92$) denotes the direction of the sweep along $U/W$, with the upward arrow indicating an increasing sweep and the downward arrow indicating a decreasing sweep. b) Finite temperature phase diagram of the ionic band insulator ($x=1.0$) obtained from HY-CTQMC. BI stands for band insulator, M for metal, and MI for Mott insulator. The red dashed curve ($U_{co}$) indicates the crossover between band insulator and metal and is extrapolated to $T=0$. The coexistence of metal and Mott insulator solutions is given by the region between $U_{c1}$ and $U_{c2}$. Three open circles at $T=0$ corresponds to DMFT-NRG calculation taken from the paper\cite{byczuk2009insulating}. Inset: Linear fit to $\tilde{A}_{1\sigma}$ in the metallic region at $\beta=92.0$. Small arrow indicates $U_{co}$. }
    \label{fig:side_by_side}
\end{figure}

The Fermi-level spectral weight is calculated from Matsubara Green's function $\tilde{A}_{1\sigma}$ = $-G_{1\sigma}(\tau=\frac{\beta}{2})/T\pi$ and plotted as a function of $U/W$ for various temperatures ($\beta=1/T$) in Fig.~\ref{fig:x1_aw0}. We will first focus on the results obtained for the lowest temperature ($\frac{1}{T}=\beta$ = 92). At low U value, the Fermi-level spectral weight $\tilde{A}_{1\sigma}$ is zero up to $\frac{U}{W}\sim  0.5$. Beyond that, it starts increasing with U and it reaches a maximum value($\sim$ 0.6) around $\frac{U}{W}\sim 1.2$. As we increase $U$ further, there is a discrete jump (first order transition, we call the transition point $U_{c2}$) and the spectral weight at the Fermi-level becomes zero. Therefore, for small values of U the system is a band insulator(BI), it crosses over to a metal (M) at an intermediate $U_{co}$, and then finally it becomes a Mott-insulator(MI) for $U$ larger than $U_{c2}$. At the same temperature ($\beta$ = 92) but now starting the simulations in the MI state and reducing $U$, the system transitions to a metallic state at $U_{c1}$, which is smaller than $U_{c2}$. The region between these two critical values (U$_{c1}$, U$_{c2}$) corresponds to the coexistence region, where metal and Mott insulator solutions simultaneously exist. As we increase the temperature, beyond $\beta$=48 the transition from M to MI turns into a crossover. 

From the data displayed in Fig.~\ref{fig:x1_aw0}, we find the crossover value (U$_{co}$) from BI to M by doing a linear fit in the region where $\tilde{A}_{1\sigma}$ grows linearly with U (shown for $\beta=92$ in the Fig.~\ref{fig:x1_aw0}). We identify the critical values U$_{c1}$ and U$_{c2}$ based on the low-frequency behaviour of the imaginary part of the self-energy since for a Mott insulator -Im $\Sigma_{1\sigma}(i\omega_n) \propto \frac{1}{\omega_n}$, while for a metal -Im $\Sigma_{1\sigma}(i\omega_n) \propto{\omega_n}$. We use the same procedure to determine critical values for each temperature and value of $x$.  Fig.~\ref{fig:x1_phase} displays the phase diagram we obtain for $x=1$. As we increase the temperature, the metallic region, which is bounded by two insulators, increases in size, while the band insulating region decreases. The coexistence region between the metal and the Mott insulator also decreases and finally disappears at $\beta$=48. In the metallic phase, the Matsubara self-energy at low frequency has a linear behaviour, which indicates the Fermi-liquid nature of the metal. Calculations at very low temperatures are numerically expensive. Based on the available data, we cannot conclusively determine whether a metallic phase exists at zero temperature by extrapolating the critical values, as the behavior may change in this regime. However, a simple extrapolation (see Fig.~\ref{fig:x1_phase}), together with NRG results from Ref.~\cite{byczuk2009insulating}, suggests the presence of a finite coexistence region rather than a single critical point as observed in the weak coupling perturbative method\cite{garg2006can}. 

\subsubsection{Covalent insulator ($x$=0):}

\begin{figure}[h!]
    \centering
    \begin{subfigure}[b]{0.46\textwidth}
        \centering
        \includegraphics[scale=0.59,trim={0 0 0 1}, 
        clip]{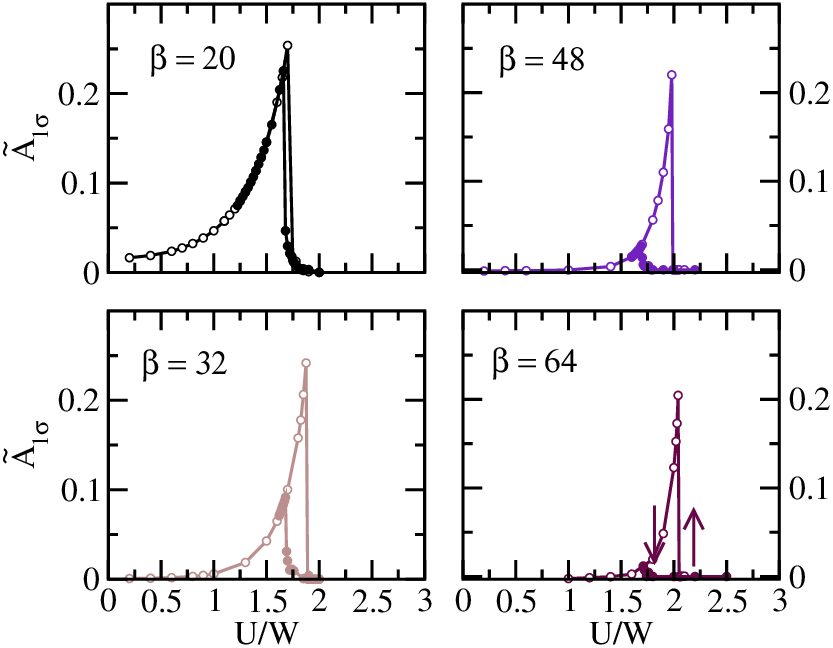}
        \caption{}
        \label{fig:x0_aw0}
    \end{subfigure}
    \hfill
    \begin{subfigure}[b]{0.46\textwidth}
        \centering
        \includegraphics[scale=0.5,trim={0 0 0 1}, 
        clip]{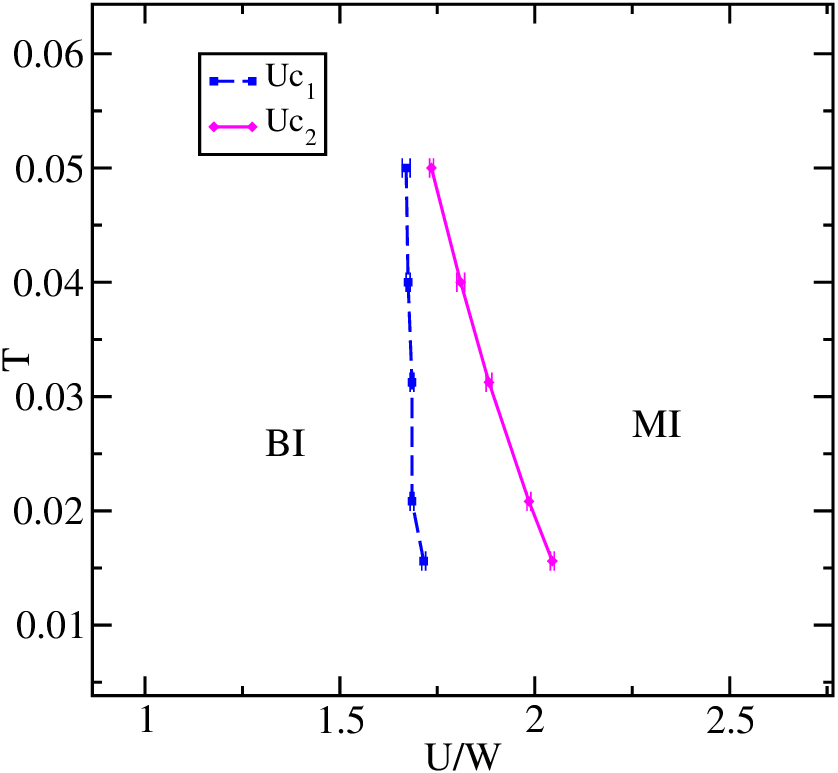}
        \caption{}
        \label{fig:x0_phase}
    \end{subfigure}
    \caption{(Color online) a) Fermi-level spectral weight as a function of $\frac{U}{W}$ for different $\beta$ values obtained from HY-CTQMC for $x$=0.0. Down-arrow corresponds to increasing U, up-arrow corresponds to decreasing U. b) Finite temperature phase diagram of the covalent insulator ($x$=0.0) on the temperature
  $T$ vs $U$ plane. BI stands for band insulator and MI for Mott insulator.}
    \label{fig:side_by_side_2}
\end{figure}

In the following, we will discuss the results obtained from HY-CTQMC. Similarly to the IHM we find that for increasing U the system goes through
a first-order transition to a MI phase with a region of phase coexistence between U$_{c2}$ and U$_{c1}$ values. However, the behavior of $\tilde{A}_{1\sigma}$ for the CI ($x$ = 0)  is completely different than for the IHM case ($x$ = 1). For example, even though both insulators have the same bandwidth, $\tilde{A}_{1\sigma}$ is zero up to a large value of U ($\frac{U}{W}\sim 1.5$ for $\beta=64$), i.e., BI phase in CI persists up to large U values. Then, $\tilde{A}_{1\sigma}$ increases quite sharply  with respect to $U$, and it is finite in a narrower range of U values in comparison with the IHM. While decreasing U, the Fermi-level spectral weight increases near Mott insulator to band insulator transition at a lesser $U=U_{c1}<U_{c2}$ and thus forming a hysteresis. This indicates the pseudo-gap nature of density of states near the transition. This is distinct from a real metallic solution. It is due to the thermal broadening of the quasi-particle peaks (square-root singularity) near the gap edges in the density of states. It is easy to verify that the appearance of the peak is due to the thermal broadening by observing the decrease in the peak height of the Fermi level spectral function as a function of the decreasing temperature. Pseudo-gap nature of the density of states, together with the thermal broadening is the reason for the peak in the spectral weight near the transition. Such a pseudo-gap density of states near the transition has been observed in the previous HY-CTQMC calculations of Sentef. et al., in the real-frequency data\cite{sentef2009correlations}.

We extract the critical values at each temperature using the procedure mentioned earlier and plot them in Fig.~\ref{fig:x0_phase}. We observe a BI phase for a broad range of U values. The coexistence region (U$_{c2}$, U$_{c1}$) between BI and MI decreases as we increase temperature. The critical values obtained from HY-CTQMC at low temperature confirm the absence of a metallic phase in the CI at zero temperature. 

\subsubsection{Equal ratio of ionicity and covalency ($x$=0.5):}
The mixing parameter $x=0.5$ corresponds to the case where ionicity and covalency are present in equal proportions. Before proceeding to the finite temperature numerical data, let's analyse the spectra at zero temperature. The structure of non-interacting Hamiltonian $\mathbf{{H}_{\sigma}(k,\mathnormal{x})}$ for $x$=0.5 is given by,
\begin{equation*}
\qquad\qquad\qquad\mathbf{H}_{\sigma}(k) = \frac{V+\epsilon_{k\sigma}}{2}
\left( \begin{array}{cc}
1 & 1\\
1 & -1
\end{array} \right)\,.
\label{eq:matrix1}
\end{equation*}
The excitation spectrum in the non-interacting case is gapless with $E_k$ = $\pm(\epsilon_k+V)/\sqrt{2}$. 
Since chemical potential $\mu=U/2$ and self-energy are zero for $U$=0, the expression for non-interacting DOS at the Fermi level $A_{1\sigma}(0)=-\frac{1}{\pi} \mathrm{Im} G_{1\sigma}(0)$ is given by,
\begin{equation*}
\qquad\qquad\qquad A_{1\sigma}(0) = \displaystyle\frac{1}{\pi} \int 
\frac{d\epsilon \,\rho_0(\epsilon)\,\eta}{\eta^2 + \frac{(\epsilon+V)^2}{2}} \,.
\end{equation*}
In the limit $\eta \rightarrow 0^+$, we get 
\begin{equation}
\begin{array}{c}
\displaystyle A_{1\sigma}(0) = \int d\epsilon\, \rho_0(\epsilon)\,
  \delta\!\left(\frac{\epsilon+V}{\sqrt{2}}\right) 
  = \rho_0(-V)\sqrt{2}\,, \\[2ex]
= \displaystyle \frac{2\sqrt{2}}{\pi D}\,
  \sqrt{1-\left(\frac{-V}{D}\right)^2}\,.
\end{array}
\label{eq:dos.at.0.5}
\end{equation}
Thus, as it corresponds to a gapless spectrum, the DOS at the Fermi-level is finite in the non-interacting case. To understand if the non-interacting metallic state survives at finite $U$, we calculate the DOS at Fermi level within the Fermi-liquid assumption. It is given by  
\begin{equation}
\begin{array}{c}
 \displaystyle A_{1\sigma}(0) = \int d\epsilon\, \rho_0(\epsilon)\,
  \delta\!\left(
    \sqrt{\frac{(\epsilon+V)^2}{4} +
    \left(\mu - \mathrm{Re}\,\Sigma_{1\sigma}(0) - \frac{\epsilon+V}{2}\right)^2}
  \right)\,.
\end{array}
\end{equation}
which is finite only if $|V| < D$ and $\frac{U}{2}-\mathrm{Re}{\Sigma}_{1\sigma}(0) = 0$. Consider
 a weakly interacting system where $U \rightarrow$ 0$^{+}$. Then,  $\frac{U}{2}-\mathrm{Re}{\Sigma}_{1\sigma}(0) \neq$ 0 since $\mathrm{Re}\Sigma_{1,\sigma}(0) \approx U n_{1\bar{\sigma}} \neq \frac{U}{2}$ due to band $1$ corresponding to larger energies. Thus, the metallic phase exists only at $U = 0$, and the system instantly becomes gapped for even an infinitesimal $U$. Thus
  apart from the non-interacting case, we again get a band insulator,
  albeit correlated, for a range of $U$ values. With increasing  $U$,  $\frac{U}{2}-\mathrm{Re}{\Sigma}_{1\sigma}(0)$ decreases, since $n_{1,\sigma} \rightarrow 0.5$. Thus, a second metallic phase, which  is correlated,  must arise at a finite $U$ value when
   $\mu-\mathrm{Re}{\Sigma}_{1\sigma}(0) = 0$. Thus, an interaction-induced insulator sandwiched between two metallic phases emerges due to local electronic correlations at zero temperature.

\begin{figure}[htp]
    \centering
    \begin{subfigure}[b]{0.46\textwidth}
        \centering
        \includegraphics[scale=0.59,trim={0 0 0 1}, 
        clip]{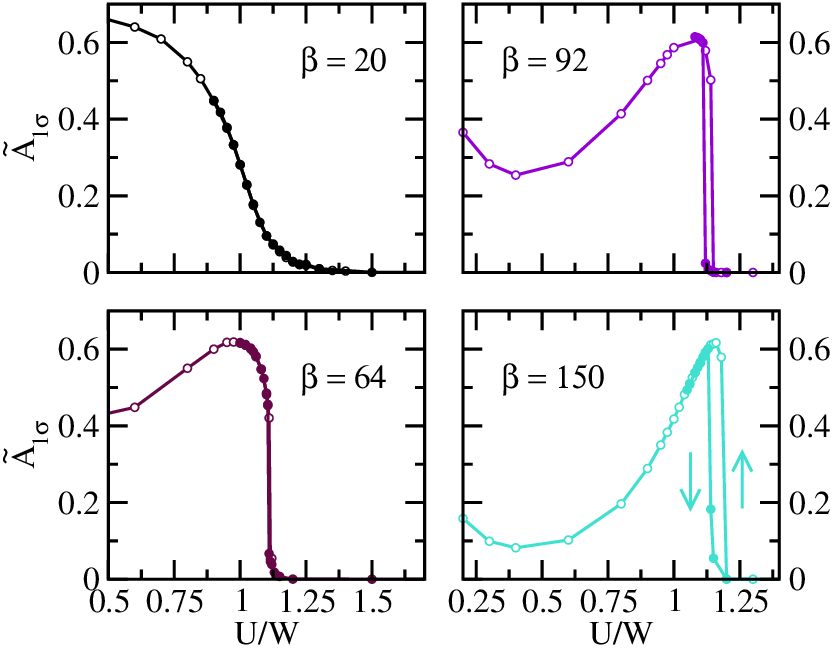}
        \caption{}
        \label{fig:x0.5_aw0}
    \end{subfigure}
    \hfill
    \begin{subfigure}[b]{0.46\textwidth}
        \centering
        \includegraphics[scale=0.49,trim={0 0 0 1}, 
        clip]{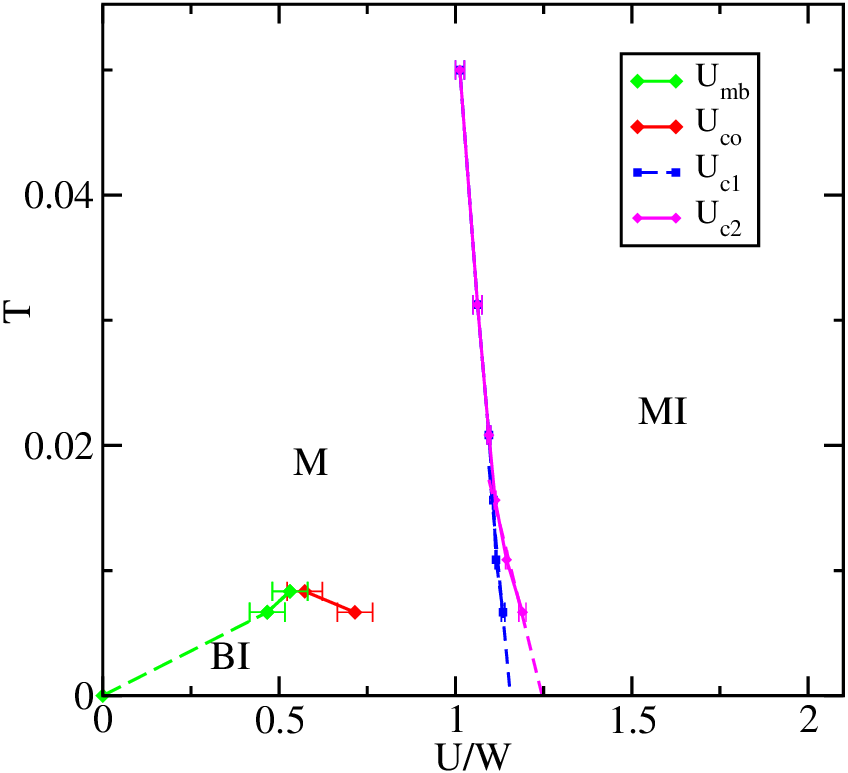}
        \caption{}
        \label{fig:x0.5_phase}
    \end{subfigure}
    \caption{(Color online) a) Fermi-level spectral weight as a function of $\frac{U}{W}$ obtained from HY-CTQMC for different $\beta$ values and $x$=0.5. b) Finite temperature phase diagram, $T$ vs $U$, for $x$=0.5 covalency. The green line ($U_{mb}$) indicates the crossover between metal and band insulator. The red curve ($U_{co}$) represents the crossover between band insulator and metal. Blue line and magenta line represents $U_{c1}$ and $U_{c2}$. The region between $U_{c1}$ and $U_{c2}$ is associated to the coexistence of metallic and Mott insulator solutions.} 
    \label{fig:side_by_side_3}
\end{figure}

Now we investigate the Fermi-level spectral weight $\tilde{A}_{1\sigma}$ as a function of U for different temperatures plotted in Fig.~\ref{fig:x0.5_aw0}. At low temperature ($\beta$=150), as we increase U, there is a minimum in $\tilde{A}_{1\sigma}$ before it reaches its highest value of $0.6$ at $\frac{U}{W}\sim 1.15$, and then the system goes to a MI state. The extrapolation of $\tilde{A}_{1\sigma}$ to the $U = 0$ axis confirms that there is a finite weight at the Fermi-level. This data suggests the existence of two metallic regions at low temperatures, one of them occurs for $\frac{U}{W}< 0.5$, and the 
other is centered at large $\frac{U}{W}\sim 1.15$. The minimum in the spectral weight points to an interaction induced band insulator emerging between these two metallic regions. As a function of increasing temperature, the minimum of $\tilde{A}_{1\sigma}$ starts filling up. The sharp transition to the MI phase and the M, MI coexistence region are also visible at $\beta > 64$ and large U values.

In Fig.~\ref{fig:x0.5_phase}, we plot the critical transition values between insulator and metal at different temperatures as a function of $\frac{U}{W}$. According to our analytical predictions, the metal at $U=0$ turns into a BI with increasing U. There is the possibility of a second metallic phase at larger U values just before the MI phase, if the condition $\mu-Re\Sigma_{1\sigma}(0)$=0 is satisfied. 
At finite temperature, we observed two metallic phases together with BI and MI phases up to $\beta\sim 100$. At very high temperature, the BI region disappears, and only M and MI regions survive, including a coexistence region for $U_{c2} < U <U_{c1}$. A linear extrapolation of metal to BI cross-over line ($U_{mb}$ shown by a dashed green line in Fig. \ref{fig:x0.5_phase}) to $T=0$ confirms the existence of a non-interacting metal. Also, a similar extrapolation of critical values of metal to MI transition validates our analytical predictions of interaction-induced BI sandwiched between two-metallic phases.

\subsubsection{General case: 0.5$<$x$<$1.0 and 0$<$x$<$0.5}
 
\begin{figure}[h!]
    \centering
    \begin{subfigure}[b]{0.49\textwidth}
        \centering
        \includegraphics[scale=0.5,trim={0 0 0 1}, 
        clip ]{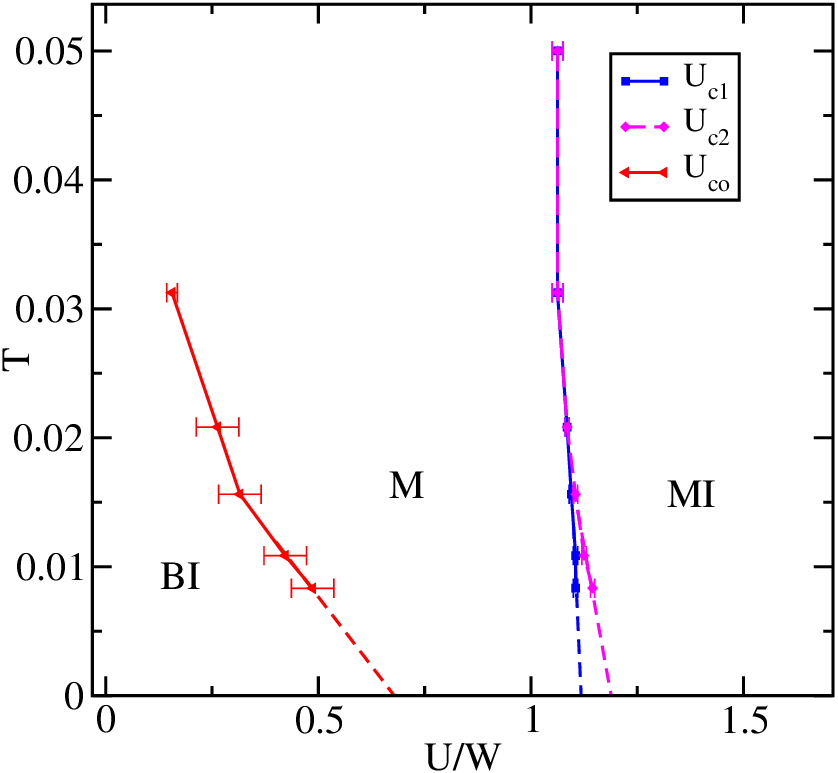 }
        \caption{}
        \label{fig:fig5.13}
    \end{subfigure}
    \hfill
    \begin{subfigure}[b]{0.49\textwidth}
        \centering
        \includegraphics[scale=0.5,trim={0 0 0 1}, 
        clip]{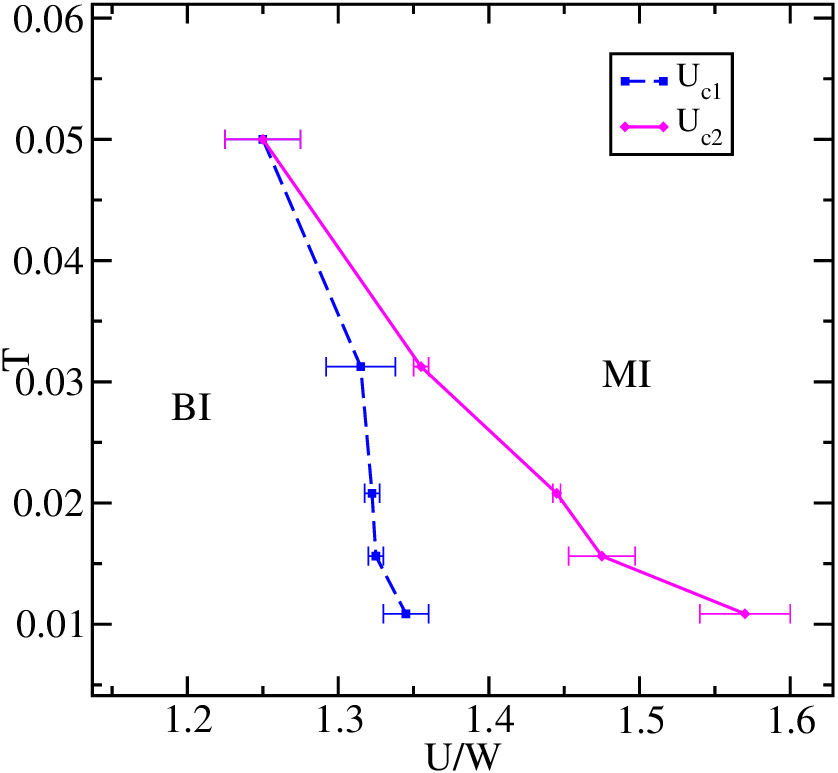 }
        \caption{}
        \label{fig:fig5.12}
    \end{subfigure}
    \caption{(Color online) Finite temperature phase diagram, $T$ vs $U$, for $x=0.75$(a) and $x=0.25$(b) covalency. The red curve ($U_{co}$) represents the crossover between band insulator and metal. Blue line and magenta line represents $U_{c1}$ and $U_{c2}$. The region between $U_{c1}$ and $U_{c2}$ is associated with the coexistence of metallic and Mott insulator solutions.}
    \label{fig:side_by_side_4}
\end{figure}

We mainly analyse two cases:  $x=0.75$ \& $x=0.25$. Fig.~\ref{fig:fig5.13} and ~\ref{fig:fig5.12} displays the $T$ vs. $\frac{U}{W}$ phase diagram of $x=0.75$ and $x$=0.25 respectively. As we decrease $x$ from 1, the metallic region between the BI and the MI increases in size, i.e., a small amount of covalency favours metallicity. Also, the coexistence region between metal and MI decreases with decreasing $x$. 
In comparison with the pure covalent case ($x$=0), the coexistence region between BI and MI for $x$=0.25 decreases and the metallic region observed for large ionicity is completely absent.   

\section{Conclusions}
\label{Model5.3}
In this work, we have investigated the role of local electronic correlations in different kinds of band insulators. Our analytical results predict the presence of a metallic point in the ionic Hubbard model (IHM), which is absent in the case of the covalent insulator (CI). When ionicity and covalency are in equal ratio, the non-interacting ground state becomes a metal, but correlations turn this non-interacting metal into a correlated band insulator. We also derive the conditions for the existence of a metallic phase for any value of the degree of ionicity. 
Our calculation using HY-CTQMC indicates the possibility of the existence of a metallic phase at zero temperature for  $0.5\le x\le1.0$, while there is no such phase for 0.0$\le x<$0.5. 
We conclude that the electronic correlations favor metallicity when the covalency is smaller than the ionicity, $x\ge 0.5$, while they favor insulating behavior when covalency is greater than ionicity, $x<0.5$. When ionicity and covalency are in equal ratio, $x$=0.5, we find the counter-intuitive result that the non-interacting ground state is a metal, which evolves to an interaction induced  band insulator (BI) with correlations. On increasing the correlation we see a transition to the metallic phase followed by a Mott insulator at higher $U$ in a discontinuous way. Our results might open new directions in the study of electronic correlations in band insulators. 
In our calculations, we have identified the crossover from a band insulator to a metal through finite density of states at the Fermi level. In the experiments, such a crossover may be identified through the evolution of photo-emission spectra, which represent the occupied density of states, as well as through transport measurements, for example, equilibrium reflectivity. Indeed, such a crossover has been observed experimentally in SrRu$_{1-x}$Ti$_x$O$_3$ by varying the doping concentration from $x=0$ to $x=1.0$ where photoemission spectra reveal the opening up of a hard gap at the Fermi-level \cite{maiti2007evolution}. Another class of experimental systems for which our findings could be relevant are 3d transition metal oxides with crystal field splitting. Although the present work addresses the issue of the region of existence of metal at zero temperature in the ionic case, some of the details, like the co-existence region and nature of the transition, require further investigations, which will be taken up in the future.

\ack

We thank CSIR and DST (India) for research funding. This work was supported by NSF DMR-1237565.
Additional support (MJ) was provided by NSF Materials Theory grant DMR1728457. Dasari acknowledges the hospitality
of the Department of Physics \& Astronomy and the Center for Computation \& Technology at Louisiana State University. Our simulations 
used an open source implementation\cite{hafermann2013efficient} of the hybridization expansion continuous-time
quantum Monte Carlo algorithm\cite{werner2006continuous}, the ALPS\cite{bauer2011alps} and the TRIQS\cite{seth2016triqs} libraries.
The CTQMC simulations were conducted on the computational resources provided by Louisiana 
Optical Network Initiative (LONI), HPC@LSU computing and National Supercomputing Mission(PARAM Yukti), India.

\section*{References}
\bibliographystyle{unsrt}
\bibliography{reference.bib}
\end{document}